# Co-NMR Knight Shift of $Na_xCoO_2{\cdot}yH_2O$ Studied in Both Superconducting Regions of the $T_c$-$\nu_{Q3}$ Phase Diagram Divided by the Nonsuperconducting Phase

Yoshiaki Kobayashi, Taketo Moyoshi, Mai Yokoi and Masatoshi Sato

*Department of Physics, Division of Material Science, Nagoya University, Furo-cho, Chikusa-ku, Nagoya 464-8602*

In the temperature ($T$)-$\nu_{Q3}$ phase diagram of $Na_xCoO_2{\cdot}yH_2O$, there exist two superconducting regions of $\nu_{Q3}$ separated by the nonsuperconducting region, where $\nu_{Q3}$ is usually estimated from the peak position of the $^{59}$Co-NQR spectra of the $\pm5/2\leftrightarrow\pm7/2$ transition and well-approximated here as $\nu_{Q3}$~$3\nu_Q$, $\nu_Q$ being the interaction energy between the nuclear quadrupole moment and the electric field gradient. We have carried out measurements of the $^{59}$Co-NMR Knight shift ($K$) for a single crystal in the higher-$\nu_{Q3}$ superconducting phase and found that $K$ begins to decrease with decreasing $T$ at $T_c$ for both magnetic field directions parallel and perpendicular to $CoO_2$-planes. The result indicates together with the previous ones that the superconducting pairs are in the spin-singlet state in both phases, excluding the possibility of the spin-triplet superconductivity in this phase diagram. The superconductivity of this system spreads over the wide $\nu_{Q3}$ regions, but is suppressed in the narrow region located at the middle point of the region possibly due to charge instability.

$Na_{0.3}CoO_2{\cdot}1.3H_2O$ with the superconducting transition temperature $T_c$ of ~4.5 K has the two-dimensional triangular lattice of Co atoms.[1)] Due to this structural characteristic, not only this system and but also the mother system $Na_xCoO_2$ have attracted much attention in relation to a possible geometrical frustration of the electron spins.[2)] To identify the mechanism of the superconducting pairing, we have clarified various kinds of their physical properties: The $T_c$ value has been found to depend on the $^{59}$Co nuclear quadrupole frequency $\nu_Q$, as summarized in the $T$–$\nu_{Q3}$ phase diagram shown in Fig. 1, $\nu_{Q3}$ (~$3\nu_Q$) being the frequency corresponding to $\pm7/2\leftrightarrow\pm5/2$ transition of the nuclear spin. The superconducting phase is divided into two regions by the nonsuperconducting phase, which can be understood by the appearance of a charge disproportionate state.[3)]

Because the increase of the crystal field $q$ at Co sites, or the $\nu_{Q3}$-increase induces the upper shift of the $e_g$'-orbital band relatively to the $a_{1g}$ one, one might expect that six hole pockets of the $e_g$'-orbital band might be expected to appear with increasing $\nu_{Q3}$ in the Fermi surface near the K points in the reciprocal space.[4)] If the hole pockets really appear, the ferromagnetic fluctuations with the wave vector of small $q$ is expected to become strong and a triplet pairing state may become possible.[5)] However, at least in the smaller-$\nu_{Q3}$ superconducting phase of the phase diagram, we have shown in previous papers that the Cooper pairs are in the singlet state, observing the suppression of the $^{59}$Co-NMR Knight shifts by the superconductivity for both magnetic field directions parallel and perpendicular to the $CoO_2$-plane.[6-8)] It is consistent with results of photoemission experiments, which have observed only hole-like Fermi surface around Γ point from the $a_{1g}$-orbital band, and never observed the $e_g$' hole pockets in the $Na_{0.3}CoO_2{\cdot}1.3H_2O$.[9)]

In the present paper, we mainly report experimental results on Co-NMR Knight shifts of $Na_{0.3}CoO_2{\cdot}1.3H_2O$ for both the magnetic field directions in the larger-$\nu_{Q3}$ superconducting phase.

$^{59}$Co-NMR and NQR measurements have been carried out using conventional coherent pulsed-NMR methods. The $Na_{0.7}CoO_2$ single crystals were prepared by the floating zone method[10)] and then Na de-intercalation and $H_2O$ intercalation were carried out as described before.[3)] The obtained single crystals were confirmed to have the $\nu_{Q3}$-values of ~12.7 MHz by NQR measurements, indicating that it is in the large-$\nu_{Q3}$ superconducting regions of the $T$-$\nu_{Q3}$ phase diagram shown in Fig. 1, where the inset shows the specific heat $C$ divided by $T$ against $T$ for a superconducting single crystal used in the present NMR (NQR) studies, indicating the bulk nature of the superconductivity with $T_c$ = ~4.7 K.

The top panel of Fig. 2 shows the ($-1/2\leftrightarrow1/2$)

transition line of the $^{59}$Co-NMR spectra for the single crystal of $Na_{0.3}CoO_2{\cdot}1.3H_2O$ with the $\nu_{Q3}$-value of ~12.7 MHz for the field ***H*** (= 1.1570 T) parallel to the *c*-axis (perpendicular to $CoO_2$ plane). In the middle panel of Fig. 2, the center line is shown with enlarged scales at several temperatures between 2 K and 6 K, indicating that the center line position is nearly *T*-independent below ~10 K down to $T_c$ (~3.5 K), and as *T* decreases, the spectra begin to shift to the lower frequency side at $T_c$. The bottom panel of Fig. 2 shows the (-1/2 ↔ 1/2) transition lines of the $^{23}$Na-NMR spectra observed at three temperatures. The line position is nearly *T*-independent in this *T* region through $T_c$, indicating that the frequency shifts of the $^{59}$Co-NMR spectra observed below $T_c$ is not due to the superconducting diamagnetism but due to the change of the magnetic susceptibility of Co spins $\chi_{spin}$. If the contribution of the superconducting diamagnetism were large, the resonance frequency shifts of Co and Na nuclei were expected to have nearly equal magnitudes, because the gyromagnetic ratios of these nuclei are nearly equal.

In Fig. 3, the shifts of the $^{59}$Co- and $^{23}$Na-NMR resonance frequency, $f_{res}(T) - f_{res}(4K)$ and the $^{59}$Co-NMR Knight shift (in the inset) for ***H*** // ***c***, $^{59}K_c$ are plotted against *T*. The absolute values of $^{59}K_c$ for ***H*** // ***c*** above $T_c$ are estimated from the seven line-positions of $^{59}$Co-NMR spectra in ***H*** (~6 T) // ***c*** and those below $T_c$ are deduced from $f_{res}(T) - f_{res}(4K)$ and the *K*-value at $T_c$. Because the *T*-dependence of $^{59}K_c$ between $T_c$ and ~150 K indicates that the hyperfine coupling constant between nuclear and electron spins is positive, $\chi_{spin}$ is found to be suppressed in the superconducting state for ***H*** perpendicular to the $CoO_2$ planes.

Next, we present the results of the $^{59}$Co-NMR Knight shift obtained for the same sample with *H*= 3.1188 T applied within the $CoO_2$ plane. The top panel of Fig. 4 shows the observed spectra of the central transition at temperatures between 10 K and ~2 K. The two peak structure of these lines indicates that the angle between the *a*-axis of the $CoO_2$ plane and the magnetic field direction distributes. Then, the peak positions of the resonance frequencies correspond to the magnetic field directions parallel to two in-plane principal axes of the electrical field gradient at Co-sites. The line begins to shift to the lower frequency side at $T_c$ with decreasing *T*, while it stays at the nearly *T* independent position above $T_c$. These results indicates that the Knight shift for ***H*** // *ab*, $^{59}K_{ab}$, is suppressed in the superconducting state. In the bottom panel of Fig. 4, the shift of the resonance frequencies $f_{res}(T) - f_{res}(6K)$ and $^{59}K_{ab}$ estimated from the peak position in the lower frequency side of the spectra are plotted against *T*. The contribution of the superconducting diamagnetic field is negligiblly small. The *T*-dependence of $^{59}K_{ab}$ indicates that $\chi_{spin}$ is also suppressed in the superconducting state for the field direction parallel to $CoO_2$ plane since the sign of the hyperfine coupling constant is positive as can be found from the $^{59}K_{ab}$-*T* curves above $T_c$. Thus, $\chi_{spin}$ of $Na_{0.3}CoO_2{\cdot}1.3H_2O$ in the larger-$\nu_{Q3}$ phase is suppressed in the superconducting state for both magnetic field directions, and we can conclude that the Cooper pairs in the larger-$\nu_{Q3}$ superconducting phase are in the singlet state.

In Fig. 5, the data of $\Delta K = K(T) - K(\sim T_c)$ presently obtained for both magnetic field directions ***H***//*ab*-plane and ***H***//***c*** are compared with those of the smaller-$\nu_{Q3}$ superconducting region.[9] The $\Delta^{59}K$-values are fitted by the Yosida function[11] for the *s*-wave superconducting order parameter to zero temperature and the values for ***H***//*ab*-plane and ***H***//***c*** are estimated to be 0.37 and 0.10 % for the larger-$\nu_{Q3}$ superconducting sample, respectively, and 0.46 and 0.15 % for the small-$\nu_{Q3}$ superconducting sample, respectively. The difference between the suppression magnitudes of the Knight shift observed for the two field directions in each superconducting phases can be understood by the anisotropy of hyperfine coupling constant.[8, 9] Considering the ambiguity of the hyperfine coupling constant, $\Delta^{59}K$ in the superconducting state for ***H***//*ab*-plane (or ***H***//***c***) seems to be $\nu_{Q3}$ independent.

Because the data in Fig. 5 indicate that $\chi_{spin}$ is suppressed in all superconducting $\nu_{Q3}$ regions and for both the field directions, we can conclude that the spin-singlet superconductivity is realized in all over the superconducting region. Thus, the spin-triplet pairing is excluded, consistently with the fact that the hole pockets of the $e_g$'-band in the Fermi surface have never been found for both hydrated and unhydrated $Na_xCoO_2$ by the angle-resolved photoemission experiments.[6] It is also consistent with our recent finding that the electronic specific heat coefficients γ above $T_c$ does not sensitively depend on $\nu_{Q3}$, because it seems to exclude the appearance of the $e_g$'-band hole pockets in the larger-$\nu_{Q3}$ regions of the phase diagram. Another recent finding of our group by neutron scattering experiments that the magnetic excitations with the wave vector ***q*** ~ 0 and energy *E*=3 meV become negligibly small as *T* decreases, also supports the results.

We think that the origin and symmetry of the superconductivity in the larger-$\nu_{Q3}$ phase are same as those in the smaller-$\nu_{Q3}$ phase, which can naturally explain the almost flat and smooth connection of the $T_c$-$\nu_{Q3}$ curves obtained in both sides of the nonsuperconducting phase. The superconducting phase spreads over the wide $\nu_{Q3}$ regions between 12.0 and 12.9 MHz, and the superconductivity is suppressed in the only narrow region of 12.5-12.6 MHz, where the charge-density-wave or the charge disproportionate state seems to take place at about 7 K and the very small amount of the magnetic moment induced inhomogeneously by this transition is frozen at lower temperatures.[3)]

In conclusion, the superconducting pairs are in the singlet state in all the superconducting $\nu_Q$ region, by studying the Co-NMR Knight shift in the superconducting state for both the magnetic field directions parallel and perpendicular to the $CoO_2$ plane. It is considered that the same types of superconductivity are taking place in both sides of the nonsuperconducting $\nu_{Q3}$ regions.

Acknowledgments –The work is supported by Grants-in-Aid for Scientific Research from the Japan Society for the Promotion of Science (JSPS) and by Grants-in-Aid on priority area from the Ministry of Education, Culture, Sports, Science and Technology.

Figure captions

Fig.1: (Color online) The $T$-$\nu_{Q3}$ phase diagram of $Na_xCoO_2{\cdot}yH_2O$. The arrow indicates the $\nu_{Q3}$ position and $T_c$ of the single crystal sample used here. The inset shows the $T$-dependence of its specific heat, indicating the bulk nature of the superconductivity.

Fig2: (Color online) Top panel shows the central transition line of $^{59}$Co-NMR spectra observed at 2.3 K. The middle panel shows the lines with the enlarged scales at several temperatures for magnetic field direction perpendicular to the $CoO_2$-planes. The bottom panel shows the central transition lines of the $^{23}$Na-NMR spectra at three temperatures. In all cases, $\boldsymbol{H}//\boldsymbol{c}$.

Fig. 3: The deviation of the $^{59}$Co and $^{23}$Na-NMR frequencies from the values at 4 K for $H$=1.1570 T. $\boldsymbol{H}//\boldsymbol{c}$. The inset shows the $T$-dependence of the $^{59}$Co-NMR Knight shift for $\boldsymbol{H} \perp ab$–plane.

Fig. 4: (Color online) The central transition lines of the $^{59}$Co-NMR spectra at several temperatures between 10 K and 1.8 K are shown as functions of the applied frequency (top) and the $T$-dependence of the deviation of the resonance frequency from the value at 6 K is also shown (bottom). In the inset, the $^{59}$Co-NMR Knight shifts are plotted against $T$. In all cases, $\boldsymbol{H}//ab$–plane.

Fig. 5: (Color online) Deviations of the Knight shifts defined as $\Delta K = K(T) - K(T \sim T_c)$ are compared for the larger- and smaller-$\nu_{Q3}$ superconducting phases. The magnetic field directions are shown in the figure.

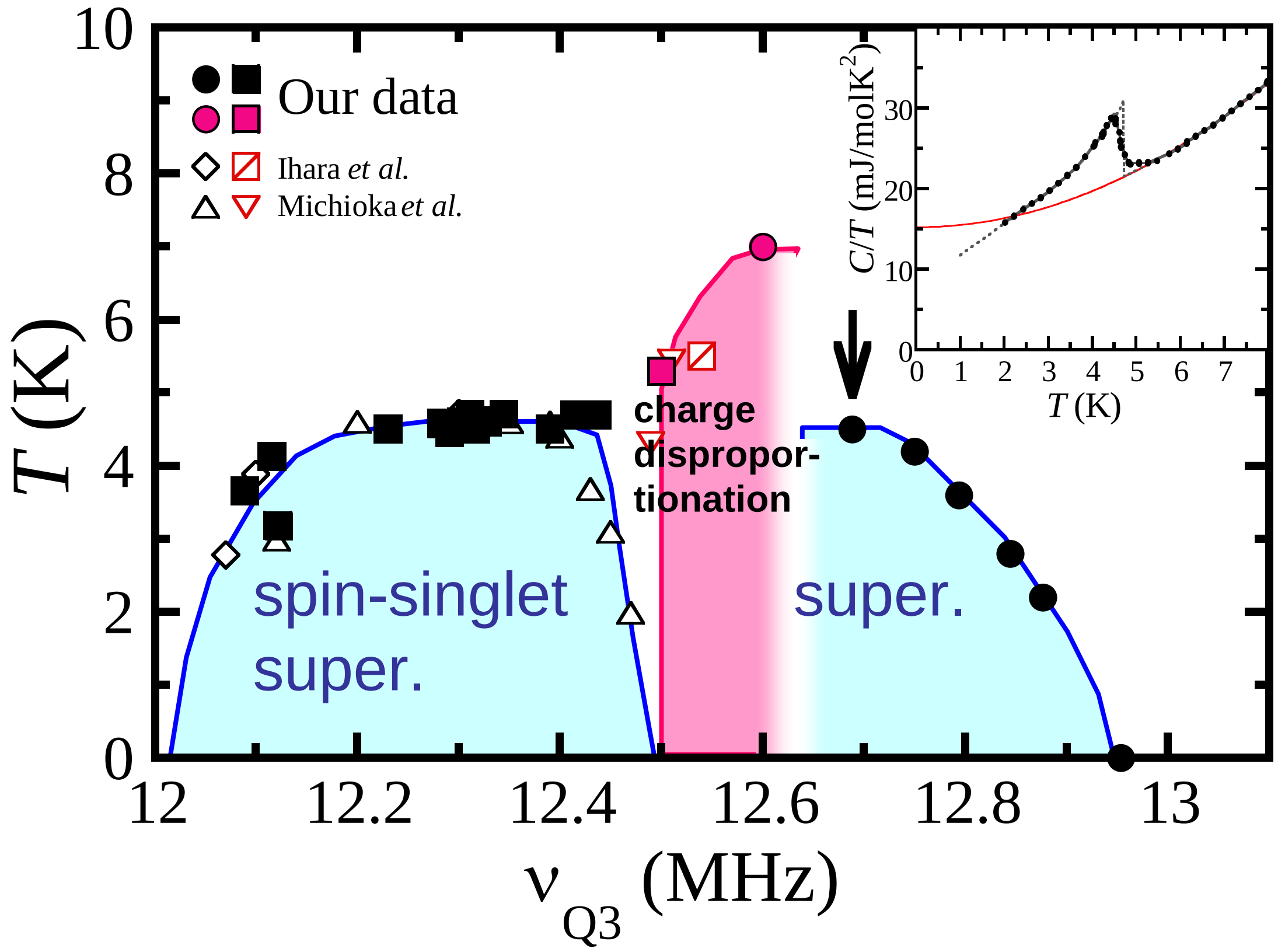

Fig. 1 Y. Kobayashi *et al.*

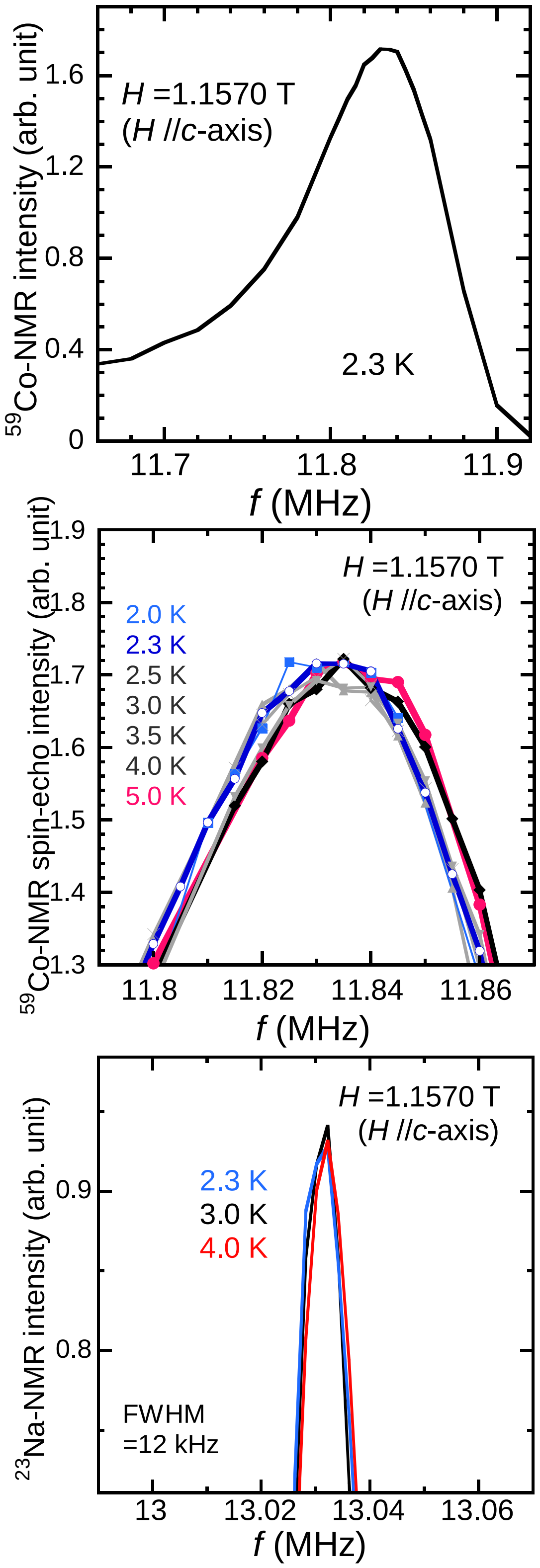

Fig. 2   Y. Kobayashi *et al.*

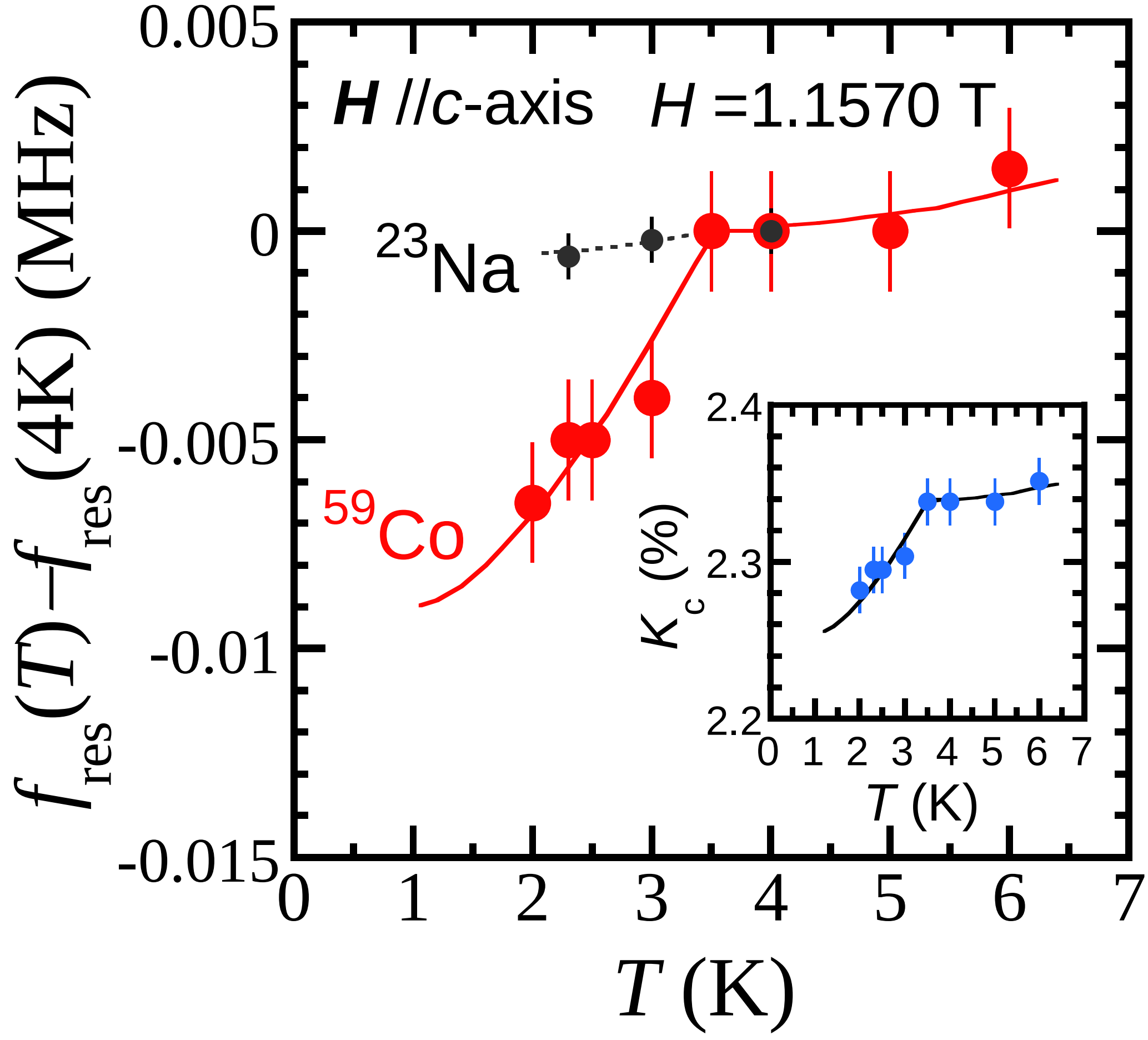

Fig. 3 Y. Kobayashi *et al.*

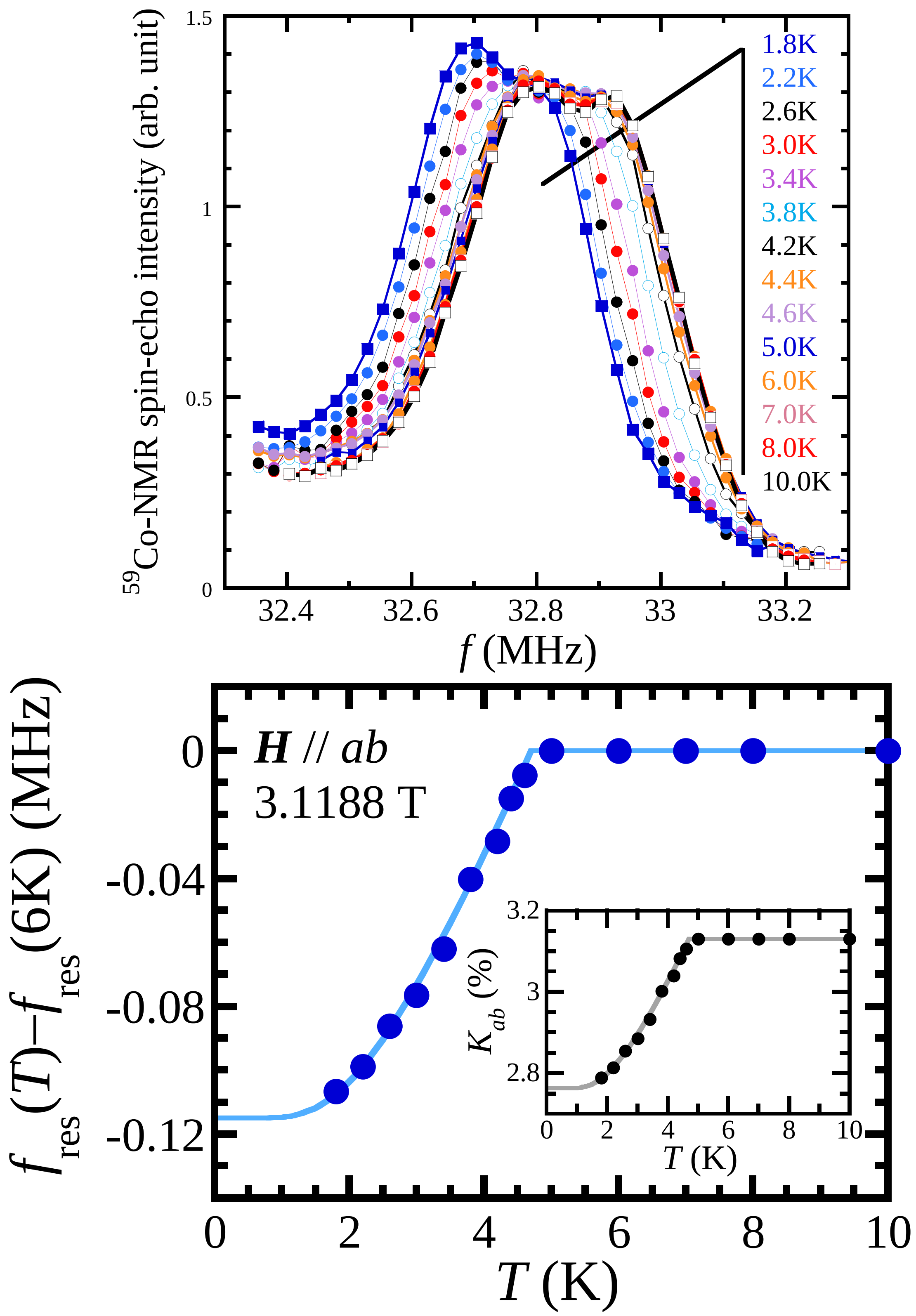

Fig. 4 Y. Kobayashi *et al.*

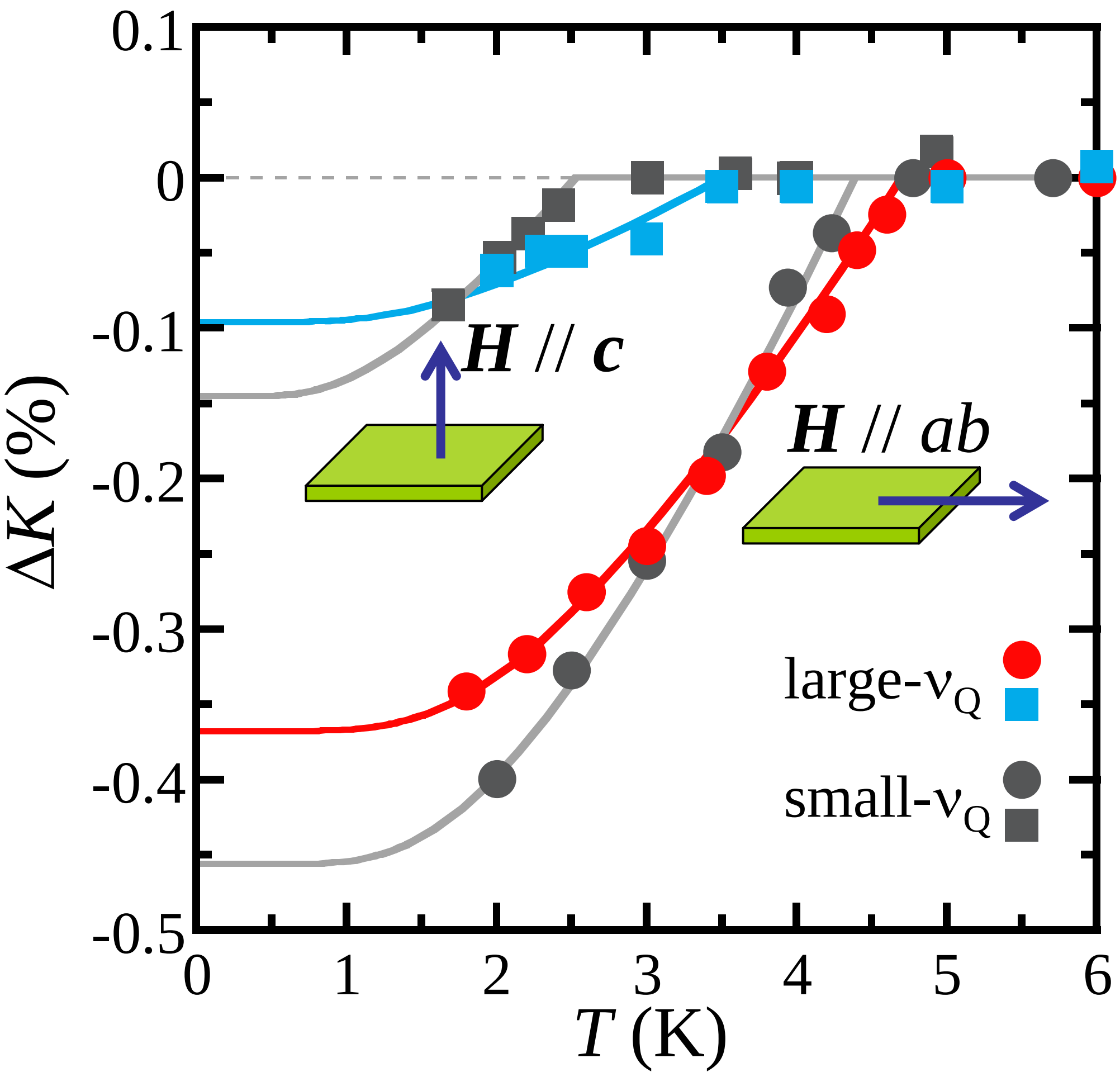

Fig. 5 Y. Kobayashi *et al.*